# Optical skyrmions in evanescent electromagnetic fields

S. Tsesses[1], E. Ostrovsky[1], K. Cohen[1], B. Gjonaj[2], N. Lindner[3], G. Bartal[1]*

[1] Andrew and Erna Viterbi Department of Electrical Engineering, Technion – Israel Institute of Technology, 3200003 Haifa, Israel.

[2] Faculty of Medical Sciences, Albanian University, Durrës st., Tirana 1000, Albania.

[3] Physics Department, Technion – Israel Institute of Technology, 3200003 Haifa, Israel.

*Correspondence to: guy@ee.technion.ac.il

**Abstract:** Topological defects play a key role in a variety of physical systems, ranging from high-energy to solid state physics. They yield fascinating emergent phenomena and serve as a bridge between the microspic and macroscopic world. A skyrmion is a unique type of topological defect, showing great promise for applications in the fields of magnetic storage and spintronics. Here, we discover and observe optical skyrmion lattices that can be easily created and controlled, while illustrating their robustness to imperfections. Optical skyrmions are experimentally demonstrated by interfering surface plasmon polaritons and are measured via phase-resolved near-field optical microscopy. This discovery could give rise to new physical phenomena involving skyrmions and exclusive to photonic systems; open up new possibilities for inducing skyrmions in material systems through light-matter interactions; and enable applications in optical information processing, transfer and storage.

**One Sentence Summary:** An optical skyrmion lattice is theoretically discovered and experimentally observed.

Topological defects are field configurations which cannot be deformed to a standard, smooth shape. They are at the core of many fascinating phenomena in hydrodynamics (*1*), aerodynamics (*2*), exotic phases of matter (*3–5*), cosmology (*6*) and optics (*7*) and, in many cases, are of importance to practical applications. The intricate dynamics of a multitude of topological defects and the efforts to control them are of key importance in high-temperature superconductivity (*8*) and topological phase transitions such as the Berezinskii–Kosterlitz–Thouless transition (*9*).

A unique type of topological defect is a skyrmion (*10*), a topologically stable configuration of a three-component vector field in two dimensions. Skyrmions were initially found theoretically in elementary particles and were since demonstrated in Bose-Einstein condensates (*11*), nematic liquid crystals (*12*) and, most famously, as a phase transition in chiral magnets (*13*, *14*). The skyrmion lattice phase and single magnetic skyrmions (*15*, *16*) are currently considered a promising route towards high-density magnetic information storage and transfer (*17*, *18*), as they are very robust to material defects and can be driven by low applied currents (*15*, *19*, *20*). A skyrmion may take on various shapes, which are all topologically equivalent. Bloch-type (*14*) and Néel-type (*21*) skyrmions exhibit a smoothly varying field configuration, with the derivatives of the vector field spread out in space. In bubble-type skyrmions (aka bubbles) (*22*, *23*), the variations of the vector field are confined to line-like areas, known as topological domain walls, which separate between two domains in which the field vectors are opposite.

In optics, topological phenomena have been a source of avid research in the past decade. Since the first observation of photonic topological insulators (*24*), optical topological phenomena have been rigorously studied, both theoretically and experimentally (*25–28*), with applications such as topologically-protected lasing (*29–31*). In fact, topological defects in optics were first extensively explored via phase and polarization singularities, both in free-space propagating light (*32–34*) and two-dimensionally confined light (*35–37*). However, only recently has there been any experimental investigation of optical topological domain walls (*38*), while the field of *optical skyrmions* has so far remained untapped.

Here, we predict and experimentally demonstrate an optical skyrmion lattice, formed by interference of guided electromagnetic waves. We show how the topological domain walls of such optical skyrmions can be continuously tuned, changing their spatial structure from bubbles to fully spread Néel-type skyrmions. We further illustrate the robustness of optical skyrmions to imperfections and generate them using low-loss surface plasmon polaritons, measured by phase-resolved near-field optical microscopy. This discovery could propel realizations and applications of skyrmions in new physical platforms: using light matter interactions, optical skyrmions could induce skyrmionic configurations in a wide variety of material, atomic and molecular systems. Furthermore, by utilizing the numerous nonlinearities known in optics, new physical phenomena involving skyrmions can now be investigated.

**Formation of optical skyrmion lattices**

The configuration of a three component real vector field on a two dimensional space provides a smooth mapping of that space to the unit sphere. The topological invariant which identifies skyrmions counts the number of times the field configuration covers the entire sphere. This topological invariant, which we denote by $S$, is known as the skyrmion number and takes integer values. For the skyrmion to be topologically robust, the space on which it is defined cannot have a boundary. This condition is indeed satisfied by a periodic field configuration, in which a skyrmion is obtained in each unit cell of a lattice. For such a "skyrmion lattice" configuration, the skyrmion number can be written in an integral form:

$$S = \frac{1}{4\pi} \int_A s \, dA \qquad (1)$$

Where the area $A$ covers one unit cell of the lattice and $s = \vec{e} \cdot \left[ \left( \partial \vec{e} / \partial x \right) \times \left( \partial \vec{e} / \partial y \right) \right]$ is the skyrmion number density; $\vec{e}$ is a real, normalized, three-component field; and $x, y$ are directions in the two dimensional plane. The skyrmion number $S$, being an integer, is robust to deformations of the field $\vec{e}$ as long as $\vec{e}$ remains non-singular and maintains the periodicity of the lattice.

Interestingly, we find that the electric field vector of electromagnetic waves can be structured so as to meet all the requirements needed to create a skyrmion lattice, similarly to the magnetization vector of the skyrmion lattices in chiral magnets. Being a three-dimensional field in a two-dimensional space, optical skyrmions must be formed by electromagnetic waves confined to two dimensions, such as in guided waves. A more thorough explanation as to why free-space electromagnetic waves do not exhibit skyrmions can be found in the Supplementary.

Consider an electric field comprised of six interfering Transverse Magnetic (TM) guided waves with equal amplitudes, freely propagating in the transverse ($x-y$) plane and evanescently decaying in the axial direction ($z$). The six waves are directed towards each other in the transverse plane and possess transverse wave vectors of similar magnitude, such that they create three standing waves in $0°$, $60°$ and $120°$. The axial (out-of-plane) field component in the frequency domain can therefore be expressed as a sum of three cosine functions:

$$E_z^{(\omega)} = E_0 e^{-|k_z|z} \sum_{\theta=-\frac{\pi}{3},0,\frac{\pi}{3}} \cos\left( k_\parallel \left[ \cos(\theta) x + \sin(\theta) y \right] \right) \qquad (2)$$

Where $E_0$ is a real normalization constant and $k_\parallel, k_z$ are the transverse and axial components of the wave vector ($k_\parallel$ is real and $k_z$ is imaginary), such that $k_\parallel^2 + k_z^2 = k_0^2$ ($k_0$ is the free-space wavenumber). The transverse (in-plane) electric field components in the frequency domain can be readily derived from Maxwell's equations:

$$\begin{pmatrix} E_x^{(\omega)} \\ E_y^{(\omega)} \end{pmatrix} = -E_0 \frac{|k_z|}{k_\parallel} e^{-|k_z|z} \sum_{\theta=-\frac{\pi}{3},0,\frac{\pi}{3}} \begin{pmatrix} \cos(\theta) \\ \sin(\theta) \end{pmatrix} \sin\left( k_\parallel \left[ \cos(\theta) x + \sin(\theta) y \right] \right) \qquad (3)$$

Namely, for waves evanescently decaying in one dimension, all electric field components are entirely real (up to a global phase), allowing to define a real unit vector $\vec{e} = \vec{E}/|\vec{E}|$ associated with the electric field and enabling a description of the field configuration as an optical skyrmion lattice.

This is a direct consequence of the phase added in the spin-momentum locking process of evanescent electromagnetic fields (39) and hence does not hold for propagating waves (see Supplementary).

Figure 1 depicts the electric field described by Eqs. (2,3), for $|k_z| \approx |k_\parallel| > k_0$. The axial electric field has the form of a hexagonally-symmetric lattice (Fig. 1A). The transverse field follows the same symmetry, yet possesses pronounced polarization singularities, which are expressed by zero-amplitude points at the center of each lattice site (Fig. 1B), at which the field direction is ill-defined (Fig. 1C). The normalized three-dimensional electric field (Fig. 1D) confirms the formation of a skyrmion lattice – each lattice site exhibits the distinct features of a Néel-type skyrmion (21), with their calculated skyrmion number, using Eq. (1), being $S = 1$.

Formation of optical skyrmion lattice can be represented in the momentum space as a transition from free-space propagation in 3D ($|k_\parallel| < k_0$) to guided mode propagation in 2D ($|k_\parallel| > k_0$). To rigorously describe this transition, we must expand the skyrmion number definition to complex electromagnetic fields, by defining $\vec{e}$ in Eq. (1), from here on out, as the *real* part of the local unit vector of the electric field. This new definition is completely consistent with the prior description of optical skyrmions in the ideal case, which is determined by Eqs. (2,3) in a lossless medium. Fig. 2 presents the transition, by showing the skyrmion number in a single lattice site as a function of $k_\parallel$ using the above definition for $\vec{e}$.

Figure 2 also shows the skyrmion number density contrast ($\psi = [s_{max} + s_{min}]/[s_{max} - s_{min}]$) in a single lattice site as a function of $k_\parallel$, which is a parameter providing a quantitative measure of the spatial confinement of the skyrmion density. For $k_\parallel$ slightly larger than $k_0$, the skyrmions are spatially confined (density contrast close to 1), with clear domain walls separating between two specific field states, effectively creating bubble-type skyrmions (point A). As $k_\parallel$ grows, the domain walls start to smear (point B), creating skyrmions with increasingly uniform skyrmion number density, converging to the Néel-type field formation shown in Fig. 1 (point C, density contrast of 0.5). This stems directly from the scaling factor $|k_z|/k_\parallel$, which can be tuned by varying the effective index (normalized propagation constant) of a guided mode.

**Observation of optical skyrmions**

Observing the optical skyrmion lattice predicted in Fig. 1,2 has several prerequisites: First, it requires a physical system allowing two-dimensional guided waves to propagate in six specific directions, while interfering them with carefully controlled phase differences. Second, it necessitates a measurement apparatus that will enable the phase-resolved imaging of such electromagnetic waves at a resolution far better than the optical diffraction limit. Additionally, a non-ideal physical system, e.g., one that is finite and with inherent losses, provides an excellent platform to examine the robustness of the topological properties of the optical skyrmions.

Such losses distort the unit vector $\vec{e} = \vec{E}/|\vec{E}|$ associated with the electric field, and create an unwanted phase difference between its components. However, in the regime of small losses, the configuration of the real part of the field still yields a well-defined skyrmion lattice, while being sufficiently larger than the imaginary part such that the latter is negligible (see full derivation in the Supplementary). In this regime, the skyrmion number $S$ deviates slightly from its quantized value at zero losses due to the weak breaking of the lattice translation symmetry, increasingly more so for unit cells which are farther away from the point of origin. Fig. 3 gives a quantitative connection between the amount of loss and the robustness of the skyrmion number, showing the number of unit cells exhibiting $S > 0.99$. For example, a configuration in which 49 sites exhibit $S > 0.99$ is obtained if the electromagnetic waves creating it persist for roughly 400 periods before decaying – an easily achieved goal in many photonic systems.

A physical system capable of fulfilling the above-mentioned requirements are surface plasmon polaritons (SPPs) (*40*) – electromagnetic surface waves existing at the interface between metallic and dielectric materials. SPPs only exist in TM polarization and the phase difference between SPPs propagating along different directions can be easily controlled (*41*). Furthermore, they are by design an imperfect system, due to the ohmic losses generated by the metal.

In the specific system used here (inset Fig. 4), SPPs are excited at the interface of air and gold, resulting in both a small propagation decay and a transverse wave vector just slightly larger than the free-space wavenumber ($k_\| = (1.038 + i0.003)k_0$). Circularly polarized light impinges on a specially designed, hexagonally-shaped coupling slit, exciting surface plasmon polaritons from each slit edge towards its center. The slit provides the same phase to the SPPs created by all edges, yet not exactly the same amplitude (due to a different propagation length of the SPPs generated by two of the edges and their propagation loss). This results in a distortion of the skyrmion lattice, which, together with the finite structure, helps to examine its robustness.

The experimental setup used to detect optical skyrmions is depicted in Fig. 4 – a scattering near-field scanning optical microscope (s-NSOM), which enables phase-resolved measurement of the electric field normal to the surface by means of pseudo-heterodyne interferometric detection (*42*). The ability to detect phase information is crucial, since it not only provides the full axial electric field, but also allows to perform a spatial Fourier transform (and its inverse), in order to filter out noises. The phase information, filtering ability and high spatial resolution of the measurement enable the correct extraction of the transverse field components and thus – of the skyrmion number.

The interference of the SPPs in the middle of the coupling slit creates a plasmonic skyrmion lattice with characteristics typical to bubbles, as portrayed in Fig. 5. This is a result of the relatively small transverse wave vector, leading to an axial field component 5 times larger than the transverse ones. While the real electric field is slightly distorted, as expected (Fig. 5A-C), it still shows similarity to the electric field configuration presented in Fig. 1. Fig. 5E depicts the extracted skyrmion number density at the center of the lattice, which resembles that of the bubble-type skyrmion lattice shown in the inset of Fig. 2 (point A). Calculating the skyrmion number in each lattice site, we reach the result $S = 0.997 \pm 0.058$, demonstrating the robustness of the optical skyrmions.

**Discussion**

We discovered and experimentally demonstrated an optical skyrmion lattice through the interference of electromagnetic waves. Our results are ubiquitous to any photonic system exhibiting evanescence, e.g. planar waveguide modes and waves undergoing total internal reflection (TIR). We further showed, theoretically, how the domain walls of the optical skyrmions can be tuned to exhibit different spatial distributions, i.e., from bubble-type to Néel-type skyrmions. We have also explored, theoretically and experimentally, the robustness of the topological properties of the system to losses, finiteness and other imperfections.

It is worth emphasizing that the optical skyrmions described in this work are in the *frequency domain,* i.e., formed in a physical field oscillating in time, as opposed to other skyrmions, which are field formations constant in time. Moreover, skyrmions in other physical systems are the result of an energetically favorable field configuration of a highly nonlinear system, whereas the optical skyrmions we present are entirely engineered by the boundary conditions of a linear system.

Our measurement apparatus, as other experimental systems for electromagnetic wave measurement, is fairly common, simple to operate and lack stringent fabrication, temperature or vacuum conditions needed in most other systems exhibiting skyrmions. As such, new physical effects, exclusive to optical skyrmions, could be rapidly and thoroughly investigated, particularly ones involving optical nonlinearities (temperature, Kerr-Type, etc.), e.g. using six guided waves propagating in a thin slab of a nonlinear crystal.

In addition, optical skyrmion lattices could possibly induce skyrmions when interacting with matter (e.g., cold atoms). Unlike optically-induced skyrmions in chiral magnets (*43*), which are *spontaneously* created by an intrinsic property of the material, optical skyrmions could drive *stimulated* creation of skyrmions in a material by the properties of light. This article focuses on TM polarized electromagnetic waves, which exhibited skyrmions in the electric field. Arranging Transverse Electric (TE) polarized waves in a similar way, using a dielectric waveguide or by TIR, would create skyrmions in the *magnetic* field of the electromagnetic wave, potentially inducing the landscape of a skyrmion lattice in different magnetic systems, avoiding the need to use exotic materials or complex fabrication. Finally, like the many possible applications for magnetic skyrmions, new schemes for optical information processing, storage and transfer could be designed using optical skyrmions.

**References and Notes**

1. H. Helmholtz, LXIII. On Integrals of the hydrodynamical equations, which express vortex-motion. *London, Edinburgh, Dublin Philos. Mag. J. Sci.* **33**, 485–512 (1867).
2. S. Goldstein, On the Vortex Theory of Screw Propellers. *Proc. R. Soc. London. Ser. A, Contain. Pap. a Math. Phys. Character*. **123**, 440–465 (1929).
3. L. Landau, Theory of the superfluidity of helium II. *Phys. Rev.* **60**, 356–358 (1941).
4. J. Bardeen, M. J. Stephen, Theory of the motion of vortices in superconductors. *Phys. Rev.* **140** (1965), doi:10.1103/PhysRev.140.A1197.
5. M. R. Matthews *et al.*, Vortices in a Bose-Einstein Condensate. *Phys. Rev. Lett.* **83**, 2498–2501 (1999).
6. T. W. B. Kibble, Topology of cosmic domains and strings. *J. Phys. A. Math. Gen.* **9**, 1387–1398 (1976).
7. L. Allen, M. W. Beijersbergen, R. J. C. Spreeuw, J. P. Woerdman, Orbital angular momentum of light and the transformation of Laguerre-Gaussian laser modes. *Phys. Rev. A*. **45**, 8185–8189 (1992).
8. G. Blatter, M. V. Feigel'Man, V. B. Geshkenbein, A. I. Larkin, V. M. Vinokur, Vortices in high-temperature superconductors. *Rev. Mod. Phys.* **66**, 1125–1388 (1994).
9. J. M. Kosterlitz, D. J. Thouless, Ordering, metastability and phase transitions in two-dimensional systems. *J. Phys. C Solid State Phys.* **6**, 1181–1203 (1973).
10. T. H. R. Skyrme, A unified field theory of mesons and baryons. *Nucl. Phys.* **31**, 556–569 (1962).
11. U. Al Khawaja, H. Stoof, Skyrmions in a ferromagnetic Bose–Einstein condensate. *Nature*. **411**, 918–920 (2001).
12. J. I. Fukuda, S. Žumer, Quasi-two-dimensional Skyrmion lattices in a chiral nematic liquid crystal. *Nat. Commun.* **2** (2011), doi:10.1038/ncomms1250.
13. C. Pfleiderer, a Rosch, a Neubauer, R. Georgii, Skyrmion Lattice in a Chiral Magnet. *Science (80-. ).*, 915–920 (2009).
14. X. Z. Yu *et al.*, Real-space observation of a two-dimensional skyrmion crystal. *Nature*. **465**, 901–904 (2010).
15. N. Romming *et al.*, Writing and Deleting Single Magnetic Skyrmions. *Science (80-. ).* **341**, 636–639 (2013).
16. D. Maccariello *et al.*, Electrical detection of single magnetic skyrmions in metallic multilayers at room temperature. *Nat. Nanotechnol.* **13**, 233–237 (2018).
17. A. Fert, V. Cros, J. Sampaio, Skyrmions on the track. *Nat. Nanotechnol.* **8** (2013), pp. 152–156.
18. N. Nagaosa, Y. Tokura, Topological properties and dynamics of magnetic skyrmions. *Nat. Nanotechnol.* **8** (2013), pp. 899–911.
19. X. Z. Yu *et al.*, Skyrmion flow near room temperature in an ultralow current density. *Nat.*

*Commun.* **3** (2012), doi:10.1038/ncomms1990.

20. J. Sampaio, V. Cros, S. Rohart, A. Thiaville, A. Fert, Nucleation, stability and current-induced motion of isolated magnetic skyrmions in nanostructures. *Nat. Nanotechnol.* **8**, 839–844 (2013).

21. I. Kezsmarki *et al.*, Neel-type skyrmion lattice with confined orientation in the polar magnetic semiconductor GaV4S8. *Nat. Mater.* **14**, 1116–1122 (2015).

22. C. Kooy, U. Enz, Experimental and Theoretical Study of the Domain Configuration in thin Layers of BaFe12O19. *Philips Res. Rep.* **15**, 7–29 (1960).

23. C. Moutafis, S. Komineas, J. A. C. Bland, Dynamics and switching processes for magnetic bubbles in nanoelements. *Phys. Rev. B - Condens. Matter Mater. Phys.* **79** (2009), doi:10.1103/PhysRevB.79.224429.

24. Z. Wang, Y. Chong, J. D. Joannopoulos, M. Soljačić, Observation of unidirectional backscattering-immune topological electromagnetic states. *Nature*. **461**, 772–775 (2009).

25. K. Fang, Z. Yu, S. Fan, Realizing effective magnetic field for photons by controlling the phase of dynamic modulation. *Nat. Photonics*. **6**, 782–787 (2012).

26. A. B. Khanikaev *et al.*, Photonic topological insulators. *Nat. Mater.* **12**, 233–239 (2013).

27. M. C. Rechtsman *et al.*, Photonic Floquet topological insulators. *Nature*. **496**, 196–200 (2013).

28. M. Hafezi, S. Mittal, J. Fan, A. Migdall, J. M. Taylor, Imaging topological edge states in silicon photonics. *Nat. Photonics*. **7**, 1001–1005 (2013).

29. B. Bahari *et al.*, Nonreciprocal lasing in topological cavities of arbitrary geometries. *Science (80-. ).* (2017), pp. 1–8.

30. G. Harari *et al.*, Topological insulator laser: Theory. *Science (80-. ).* (2018) (available at http://science.sciencemag.org/content/early/2018/01/31/science.aar4003.abstract).

31. M. A. Bandres *et al.*, Topological insulator laser: Experiments. *Science (80-. ).* (2018) (available at http://science.sciencemag.org/content/early/2018/01/31/science.aar4005.abstract).

32. N. B. Simpson, K. Dholakia, L. Allen, M. J. Padgett, Mechanical equivalence of spin and orbital angular momentum of light: an optical spanner. *Opt. Lett.* **22**, 52 (1997).

33. M. S. Soskin, V. Gorshkov, M. Vasnetsov, J. Malos, N. Heckenberg, Topological charge and angular momentum of light beams carrying optical vortices. *Phys. Rev. A*. **56**, 4064–4075 (1997).

34. F. Flossmann, U. T. Schwarz, M. Maier, M. R. Dennis, Polarization singularities from unfolding an optical vortex through a birefringent crystal. *Phys. Rev. Lett.* **95** (2005), doi:10.1103/PhysRevLett.95.253901.

35. Y. Gorodetski, A. Niv, V. Kleiner, E. Hasman, Observation of the spin-based plasmonic effect in nanoscale structures. *Phys. Rev. Lett.* **101** (2008), doi:10.1103/PhysRevLett.101.043903.

36. G. Spektor *et al.*, Revealing the subfemtosecond dynamics of orbital angular momentum

36. in nanoplasmonic vortices. *Science (80-. ).* **355**, 1187–1191 (2017).

37. E. Ostrovsky, K. Cohen, S. Tsesses, B. Gjonaj, G. Bartal, Nanoscale control over optical singularities. *Optica.* **5**, 283–288 (2018).

38. M. Gilles *et al.*, Polarization domain walls in optical fibres as topological bits for data transmission. *Nat. Photonics.* **11**, 102–107 (2017).

39. T. Van Mechelen, Z. Jacob, Universal spin-momentum locking of evanescent waves. *Optica.* **3**, 118–126 (2016).

40. S. A. Maier, *Plasmonics: Fundamentals and applications* (2007).

41. G. Spektor, A. David, G. Bartal, M. Orenstein, A. Hayat, Spin-patterned plasmonics: towards optical access to topological-insulator surface states. *Opt. Express.* **23**, 32759 (2015).

42. N. Ocelic, A. Huber, R. Hillenbrand, Pseudoheterodyne detection for background-free near-field spectroscopy. *Appl. Phys. Lett.* **89** (2006), doi:10.1063/1.2348781.

43. G. Berruto *et al.*, Laser-Induced Skyrmion Writing and Erasing in an Ultrafast Cryo-Lorentz Transmission Electron Microscope. *Phys. Rev. Lett.* **120**, 117201 (2018).

44. T. V Teperik, A. Archambault, F. Marquier, J. J. Greffet, Huygens-Fresnel principle for surface plasmons. *Opt. Express.* **17**, 17483–17490 (2009).

**Acknowledgments**

The Authors would like to thank S. Dolev for his help in fabrication of the measured samples. **Funding:** "Circle of Light," Israeli Centers for Research Excellence (I-CORE); Israel Science Foundation (ISF) (1802/12). **Author contributions:** S.T., B.G. and G.B. conceived the project, S.T. and K.C. patterned the samples, E.O., S.T. and K.C. performed the measurements, S.T., N.L. and G.B. performed simulations and analytical calculations. All authors took part in preparing the manuscript. **Competing interests:** The authors declare no competing interests. **Data and materials availability:** All data is available in the main text or the supplementary materials.

**Supplementary Materials**

Materials and Methods

Figures S1-S3

References (*41*, *42*, *44*)

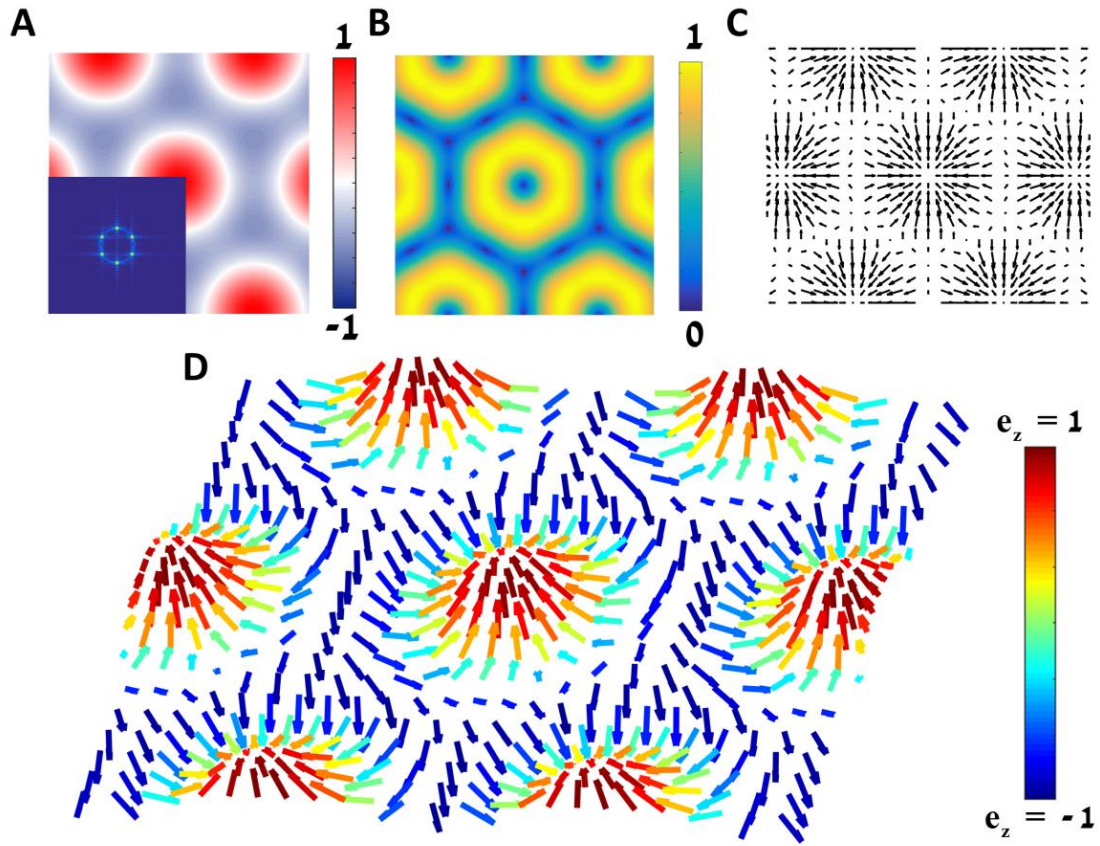

**Fig. 1. Calculated Electric field distribution of the optical skyrmion lattice.** (**A**) Axial (out-of-plane) electric field, according to Eq. 2, with the Fourier decomposition in the inset. (**B**) Amplitude of the transverse (in-plane) electric field (Eq. 4). (**C**) Vector representation of the transverse electric field, showing polarization singularities at the center of each lattice site. (**D**) Vector representation of the local unit vector of the electric field (color coded for the value of its axial component), showing that each lattice site is a Néel-type skyrmion.

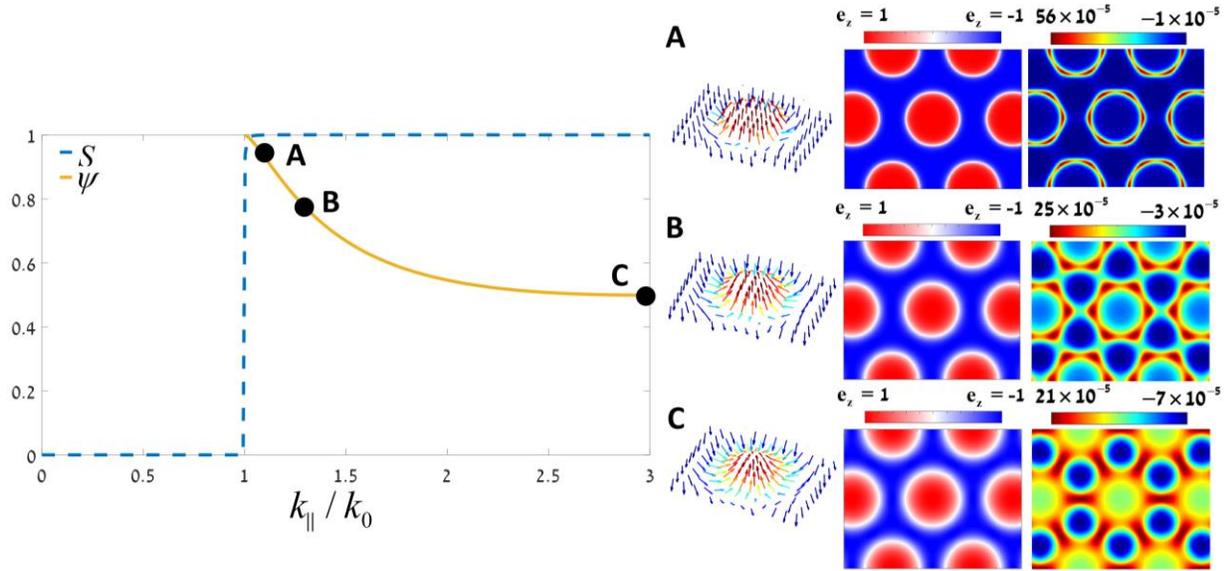

**Fig. 2. Tuning of optical topological domain walls, from bubble-type to Néel-type skyrmions.** the figure shows the calculated Skyrmion number (blue, dashed) and the skyrmion number density contrast (solid, gold) as a function of the transverse wave vector of the electromagnetic waves. Once the electromagnetic waves gain evanescence in free-space, the skyrmion number jumps to the value of 1, independent of the change in the transverse wave vector. Points **A-C** have a different skyrmion number density contrast, quantifying the transition from bubble-type to Néel-type skyrmions. Inset are the characteristics of a single lattice site at each of these points, presenting (from left to right): the local unit vector of the electric field (color coded in the same manner as Fig. 1); its axial (out-of-plane) component; and the skyrmion number density. A clear shift from very narrow domain walls separating between two field states (point **A**), to smeared domain walls (point **B**) and finally – to virtually non-local domain walls (point **C**), arises by increasing the transverse wave vector.

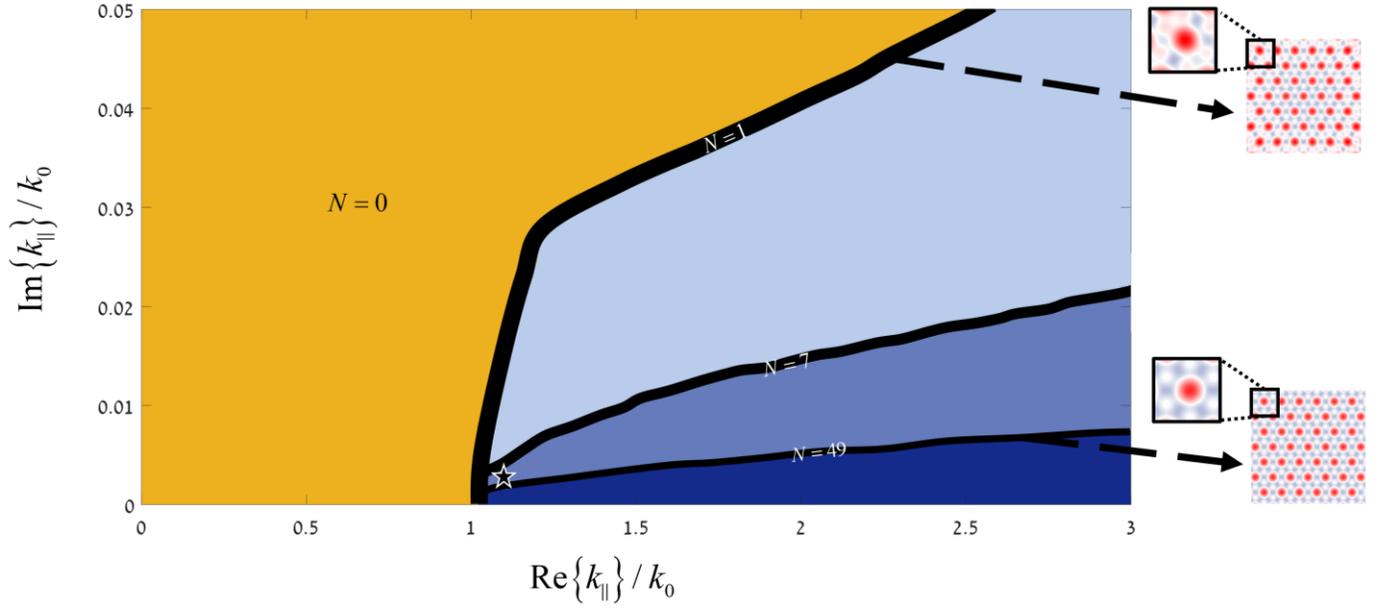

**Fig. 3. Theoretical persistence of optical skyrmions in the presence of loss.** The figure shows a contour plot of the maximal number of well-defined skyrmions in a lattice ($N$), as a function of the real and imaginary parts of the transverse wave vector. The broad black lines represent the areas where the lattice consists of $N = 1, 7, 49$ well-defined skyrmions. Insets are the amplitude distributions of an $N = 1$ lattice (top right) and an $N = 49$ lattice (bottom right), with a close-up image showing the lattice distortion in non-central sites. Clearly, in the $N = 1$ lattice, only the central lattice site is well-defined. The star represents the experimental conditions of this work ($N = 37$). These results show that by determining the number of well-defined skyrmions in a lattice, it is possible to determine the maximal amount of losses which will destroy its topological invariant. Moreover, they show that even guided modes with a relatively high amount of loss (decaying after 100 oscillations) can create skyrmion lattices.

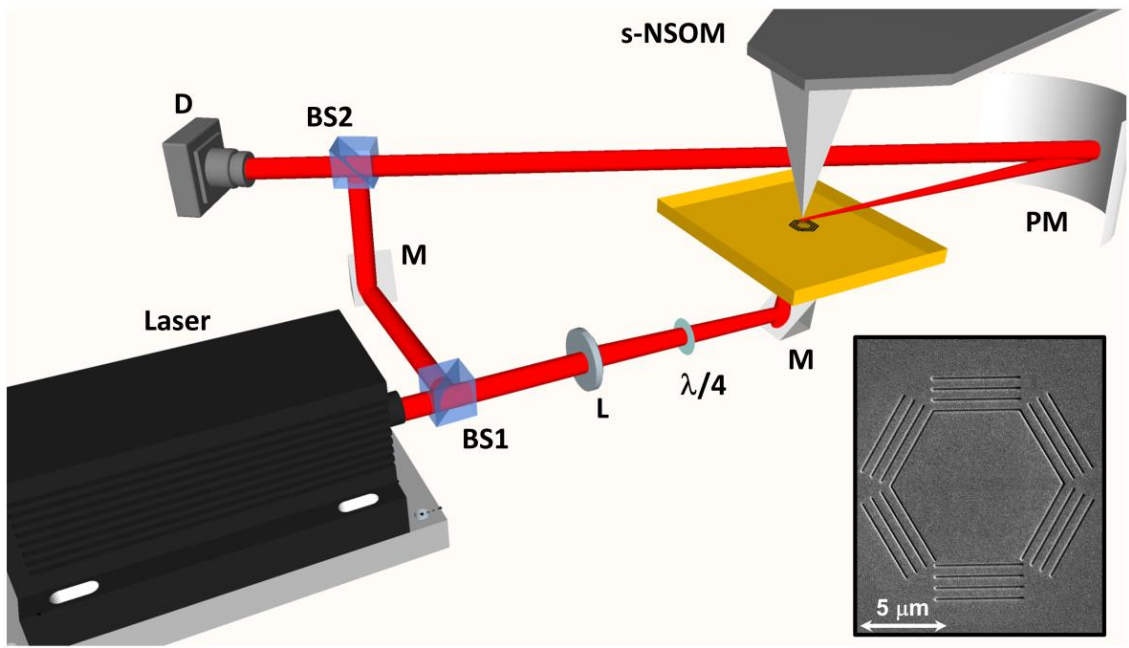

**Fig. 4. Experimental setup and sample characterization.** 660 nm laser light passes through a beam splitter (BS1), then through a weakly focusing lens (L) and a quarter-wavelength plate ($\lambda/4$). Afterwards, light impinges on the sample in circular polarization from below, using a periscope mirror (M). Surface waves are excited from the slit and propagate towards its center. The near-field signal is then scattered by the s-NSOM system to a parabolic mirror (PM). It then passes through another beam splitter (BS2) and is interfered with the incident light on the detector (D). Inset is an SEM image of the sample (with a scale bar at the bottom-left corner). The sample consists of six gratings creating a hexagon on a 200 nm Au layer, deposited on a 1 mm glass substrate. The grating periodicity corresponds to the plasmonic wavelength (636 nm) and the bottom grating is displaced by half the plasmonic wavelength, to create the necessary phase relations between the interfering waves.

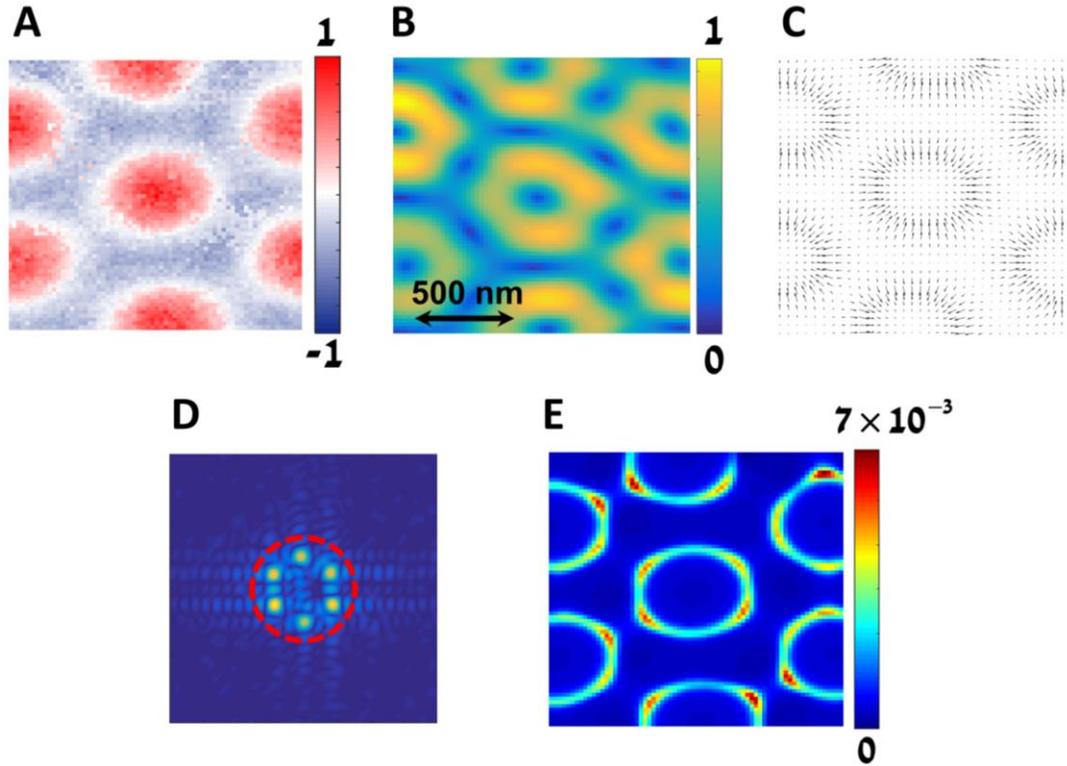

**Fig. 5. Measurement of an optical skyrmion lattice created by surface plasmon polaritons.**
(**A**) Real part of the axial (out-of-plane) electric field at the center of the sample described in Fig. 3 (without noise-filtering). (**B**) Amplitude of the transverse (in-plane) electric field. (**C**) Vector representation of the transverse electric field. (**D**) Amplitude of the Fourier decomposition of the longitudinal electric field. (**E**) Skyrmion number density map of the lattice. Scale bar is inset (**B**). The complex value of the axial electric field was measured with 20 nm resolution by the system described in Fig. 3, and measurement noise was filtered with the low-pass filter appearing in (**D**) as a red (dashed) circle. (**B,C,E**) were then extracted from the data. The measured skyrmion number of a single lattice site was $S = 0.997 \pm 0.058$, which, together with the figures, shows a good match to the theoretical prediction. Actual number of lattice sites in our sample was 37.